\documentclass[conference]{IEEEtran}

\usepackage{amsmath}
\usepackage{graphics}
\usepackage{setspace}
\usepackage{cite}
\usepackage{latexsym}
\usepackage{float}
\usepackage{epsfig}
\usepackage{multirow}
\usepackage{cite,cases,url}
\usepackage{amssymb}
\usepackage{graphicx}
\usepackage{epstopdf}
\usepackage{balance}
\usepackage{enumerate}
\usepackage{array}
\usepackage{algorithmic}
\usepackage{algorithm}
\usepackage{color}
\usepackage{soul}

\newcolumntype{L}{>{\centering\arraybackslash}m{5cm}}
\newcolumntype{K}{>{\centering\arraybackslash}m{6cm}}
\newcolumntype{P}{>{\centering\arraybackslash}m{2.3cm}}
\newcolumntype{M}{>{\raggedright\arraybackslash}m{2cm}}
\newcolumntype{N}{>{\raggedright\arraybackslash}m{2.5cm}}

\begin{document}

\title{Software Radios for Unmanned Aerial Systems}

\author{
\IEEEauthorblockN{Keith Powell$^1$, Aly Sabri Abdalla$^1$, Daniel Brennan$^1$, Vuk Marojevic$^1$, R. Michael Barts$^2$, Ashwin Panicker$^3$}
\IEEEauthorblockN{Ozgur Ozdemir$^3$, and Ismail Guvenc$^3$}
\IEEEauthorblockA
{
$^1$Department of Electrical and Computer Engineering, Mississippi State University, Mississippi State, MS\\
$^2$Wireless Research Center of North Carolina, Wake Forest, NC\\
$^3$Department of Electrical and Computer Engineering, North Carolina State University, Raleigh, NC}
{\{kp1747,asa298,dmb845,vm602\}@msstate.edu, mike.barts@wrc-nc.org, \{ampanick,oozdemi,iguvenc\}@ncsu.edu}


}
\maketitle



\begin{abstract}

As new use cases are emerging for unmanned aerial systems (UASs), advanced wireless communications technologies and systems need to be implemented and widely tested. 
This requires a flexible platform for development, deployment, testing and demonstration of wireless systems with ground and aerial nodes, enabling effective 3D mobile communications and networking. 
In this paper, we provide a comparative overview of software-defined radios (SDRs), with a specific focus on SDR hardware and software that can be used for aerial wireless experimentation and research. We discuss SDR hardware requirements, features of available SDR hardware that can be suitable for small UASs, and power measurements carried out with a subset of these SDR hardware. We also present SDR software requirements, available open-source SDR software, and calibration/benchmarking of SDR software. As a case study, we present AERPAW: Aerial Experimentation and Research Platform for Advanced Wireless, and discuss various different experiments that can be supported in that platform using SDRs, for verification/testing of future wireless innovations, protocols, and technologies. 



\end{abstract}

\IEEEpeerreviewmaketitle

\section{Introduction}
\label{sec:intro}

Unmanned aerial systems (UASs) consist of unmanned aerial vehicles (UAVs), UAV controllers, and the UAS traffic management (UTM).
The UAV controllers are radio controller (RC) units that allow a pilot to fly the UAV. 
Current regulations for standard UAV flights require a pilot to be in visual line of sight of the UAV.
It is envisaged that advanced wireless networks under the 5G umbrella, will provide the radio frequency (RF) connectivity for both command and control (C2), or control and non-payload communications, and data links, or payload communications. 
The Third Generation Partnership Project (3GPP), which standardizes global cellular communications networks, has therefore begun to evaluate critical features and develop standards for integrating UAVs into cellular networks and providing a reliable and global wireless network for UAS communications. 
The predicted growth of UAVs for commercial purposes \cite{kakar2017waveform} will be enabled by advanced wireless networking which is the foundation for the safe integration of UAVs into the national airspace (NAS). 

The robust and high-capacity communications and autonomous control  introduced by 5G wireless systems will enable many additional applications and services, including but not limited to: 1) terrestrial network data offloading and coverage extension~\cite{kumbhar2017exploiting}; 2) RF monitoring for secure and dynamic spectrum access~\cite{8453329}; 3) surveillance for safety, security, or law enforcement~\cite{7842423}; 4) Public safety and ad hoc networking for disaster relief~\cite{UABSforPSC_VTmag2016}; 5) Precision agriculture~\cite{mogili2018review}; and 6) Vehicular traffic monitoring and regulation~\cite{UAVsmartcities_CommMag2016}. While the emerging 5G communications technology and networks will be relevant for UAS, new wireless protocols and systems, or communications modes, will need to be designed for several reasons, including:
\begin{itemize}
    \item Most traditional wireless protocols and systems are optimized for terrestrial and satellite communications;
    \item UAVs need to integrate highly-reliable 
    command and control and high-throughput data links;
    \item UAVs come in many different forms and capacities, and communications systems for these need to be scalable.
\end{itemize}

Traditional radio communications for air traffic control use obsolete technology and protocols which are not suitable for modern high-capacity communications in the dynamic RF spectrum. 
Future research will cover the design, development and testing of new communications systems, or system adaptations, for supporting the proliferation of a variety of UASs that operate fully autonomously and coexist with other ground and aerial wireless network users and infrastructure. 
There exists a multitude of theoretical and simulation-based proposals for UAV communications and networking. 
On the other hand, experimental research on UAV communications is limited, due to the complex end-to-end systems that demand exhaustive testing of the different system components for public awareness and adoption into standards. 
For example, it is not practical to deploy a UAV base station (BS) that can be aloft for 30 minutes or that lacks the backhaul.

While the academia, industry, and governments have tested UAS communications and networking for certain UAS missions, they are mostly used for isolated use cases, in confined environments, and for specific missions. 
The integration of UASs into different air spaces and the cooperation, coordination and control of resources --airspace, network capacity, and UAS internal resources-- require a wide-area, software-configurability and an extensible 3D communications platform for wireless research in a real environment with experimental and production-type systems. 
In this paper, we consider \emph{AERPAW: Aerial Experimentation and Research Platform for Advanced Wireless}~\cite{aerpaw}, which is an advanced wireless testbed for UASs, deployed in a non-isolated environment. AERPAW's unique testing environment will support emerging technologies such as 4G/5G,  millimeter-wave (mmWave) systems, and massive Internet of Things (IoT) connectivity, 
to realize unforeseen broadband applications and accelerate advances of autonomous transportation systems. This paper will in particular describe the software-defined radio (SDR) development efforts from both hardware and software perspectives with a focus on UASs, taking AERPAW as a case study. 

The remainder of this paper is organized as follows. 
Sections II and III introduce the SDR use cases for aerial scenarios, hardware and software requirements, and proposed systems with initial benchmarking results. 
Section IV briefly introduces the AERPAW testbed, its research requirements, radios, experimental flow, and enabled experiments. 
Finally, Section~V derives the conclusions.



\section{Advanced Wireless Use Cases for UAVs}
\label{sec:uav}


There are various advanced wireless use cases for UAVs where SDRs can be used  to research, develop, and test new approaches. Some representative examples are as follows.

\textbf{Aerial Relay:}
An aerial relay UAV~\cite{gangula2018flying} can  transmit exchanged packets between a ground transmitter and a terrestrial base station. This scenario can provide a high signal strength at a low transmission power. This role can give a deeper insight to studies on improving cellular network security~\cite{AlyIoT20}. 

\textbf{Aerial Attacker:}
This role involves using UAVs to act as a malicious node in the wireless network, e.g. an eavesdropper or jammer, or it could be a friendly jammer~\cite{ShangV2X-19}. 
The line-of-sight communication links can enhance the effect of such attacks. Those scenarios can help with investigating and modeling aerial attacks for advancing wireless security systems.

\textbf{Aerial Repeaters:}
UAVs as secondary access points can enable experiments 
that study optimal aerial positioning and path planning of aerial nodes to enhance cellular coverage of ground cellular nodes in challenging radio environments. This can also enable various other experiments including those involving high mobility ad-hoc networks such as Vehicle-to-X. 

\textbf{Aerial RF Sensing and Spectrum Sharing:}
RF spectrum sensing by the ground and aerial nodes will be key to efficient use of RF spectrum. The UAVs may need to share spectrum with terrestrial users and employ advanced spectrum sensing and access mechanisms, leveraging the increasing amount of unlicensed and shared spectrum \cite{ShangVTM20}.

\textbf{Aerial Scouts:}
The UAV can be deployed in different environments for monitoring and sensing different environmental parameters. This role can advance different field of sensor applications such as agriculture, spatial ecology, pest detection, wildlife species tracking, search and rescue, target tracking, the monitoring of the atmosphere, chemical, biological, and natural disaster phenomena, fire prevention, flood prevention, volcanic monitoring, and pollution monitoring. Additionally, aerial scouts can be used for sensing and monitoring wireless communication links to provide more insight on the different RF parameters. This information can be used to enhance handover procedures, resource allocation, interference, and network load/offload.

\section{SDR Hardware}
\label{sec:hardware}



\begin{table*}[ht!]
\scriptsize 
\centering
\caption{Comparison of the features of different SDR hardware.} \label{Tab:SDR_Compare}
\vspace{-2mm}
\begin{tabular}{|p{1.1cm}|p{0.7cm}|p{0.7cm}|p{0.7cm}|p{1.4cm}|p{1.3cm}|p{1cm}|p{1.2cm}|p{1.2cm}|p{1.0cm}|p{0.9cm}|p{1cm}|}
\hline
\textbf{SDR Peripheral} & \textbf{Sample Rate} & \textbf{Inst. BW} & \textbf{MIMO} & \textbf{Interface}  & \textbf{Frequency Range} & \textbf{ADC/DAC bits} & \textbf{Max TX Power}  & \textbf{Max Input Power}& \textbf{Noise Figure} & \textbf{Use} & 
\textbf{DC Input}
\\ \hline 
USRP 2974 & 200 Msps & 160 MHz & 2x2 & Embedded/{1\& 10 GigE}/PCIe & 10 MHz - 6000 MHz & 14/16 & 20 dBm & 10 dBm & 5 dB -\newline7 dB & BS & 14.25 V -15.75 V
\\ \hline
USRP X3xx & 200 Msps & 160 MHz & 2x2 & {1 \& 10 GigE}/PCIe & DC - \newline6000 MHz$^{\mathrm{a}}$ & 14/16 & $>$10 dBm$^{\mathrm{b}}$& -15 dBm & 8 dB & BS & 12 V
\\ \hline
USRP N3xx & 200 Msps & 100 MHz & 2x2 / 4x4 & {1 \& 10 GigE}/PCIe & 10 MHz - 6000 MHz & 16/14 & 18 dBm$^{\mathrm{b}}$ & -15 dBm & 5.8 dB -\newline 7.5 dB & BS & 12 V
\\ \hline
USRP B20x mini & 61.44 Msps & 56 MHz & No & USB 3.0 & 70 MHz - 6000 MHz & 12/12 & $>$10 dBm & 0 dBm & $<$8 dB & UE/BS & 5 V USB
\\ \hline
USRP B200 & 61.44 Msps & 56 MHz & No & USB 3.0 & 70 MHz - 6000 MHz & 12/12 & $>$10 dBm & 0 dBm & $<$8 dB & UE/BS & 6 V
\\ \hline
USRP B210 & 61.44 Msps & 56 MHz & 2x2 & USB 3.0 & 70 MHz - 6000 MHz & 12/12 & $>$10 dBm & 0 dBm & $<$8 dB & UE/BS & 6 V 
\\ \hline
USRP E31x & 61.44 Msps & 56 MHz & 2x2 & Embedded  & 70 MHz - 6000 MHz & 12/12 & $>$10 dBm & 0 dBm & $<$8 dB & UE/BS & 5 V -\newline 15 V
\\ \hline
USRP E320 & 61.44 Msps & 56 MHz & 2x2 & Embedded/{1 \& 10 GigE} & 70 MHz - 6000 MHz & 12/12 & $>$10 dBm & -15 dBm &  $<$8 dB & UE/BS & 10 V -\newline 14 V
\\ \hline
Lime SDR & 61.44 Msps & 61.44 MHz & 2x2 & USB 3.0 & 100 kHz - 3800 MHz & 12/12 & 10 dBm$^{\mathrm{b}}$ & N/A & N/A & UE/BS & 5 V USB
\\ \hline
bladeRF 2.0 micro xA4/xA9 & 61.44 Msps & 56 MHz & 2x2 & USB 3.0 & Tx/Rx: 47/70 MHz -\newline 6000 MHz
& 12/12 & 8 dBm & N/A & N/A & UE/BS & 5 V USB
\\ \hline
\multicolumn{12}{c}{$^{\mathrm{a}}$The actual value is dependent on the installed daughter-boards. \qquad $^{\mathrm{b}}$ Max power is dependent on center frequency.}
\end{tabular}
\vspace{-2mm}
\end{table*}


\subsection{Hardware Requirements}

The following are the key requirements for the different hardware components of a modern software radio:

\textbf{High-performance and flexible RF:} Modern systems are wideband and frequency agile, requiring suitable RF components and antennas. RF front ends need to be modular for frequency agility and operation at micro and millimeter wave, with different types of antenna systems, including massive antenna arrays.

\textbf{Flexible data conversion:} To achieve frequency agility and trade performance for cost, multiple or exchangeable conversion stages are needed for effective data acquisition and generation of wideband waveforms in different frequency bands of interest in 5G and beyond. 

\textbf{High capacity processors:} Certain applications require high single-core performance, others are paralellizable to leverage parallel computing. Latency dependant applications, such as 5G development, require high clock speeds in order to operate the radios in real-time. Other applications, such as GNURadio, have the ability to scale with the number of processor threads available. Our base stations use Intel i9 processors which deliver a high per-core performance with plenty of cores available. To make the most usage of these processors, sleep states, CPU frequency scaling, and hyperthreading are all disabled, as recommended by OpenAirInterface (OAI).

UAVs will use smaller systems---Intel NUC8 or NUC10---which are low power consuming and lightweight computers in a small form factor package. These devices use the Intel i7 U-series processors with up to 64 GB of RAM and several USB 3.0 ports. This allows for high performance with USB capable SDRs, while minimizing the power consumption. 

\textbf{High-speed interfacing for high performance data links:} The data needs to be efficiently piped between the radio and the computer, leveraging operating system services or hardware specific tools. 
Modern broadband wireless protocols implemented as SDRs using a host computer, require high-throughput and low-latency interfacing.


\subsection{SDR Hardware}
The popularity of software radios, due in part to advances in computing and the emergence of free open-open source software libraries over the past decades led to the emergence of a variety of software radio hardware of different form factors, performance figures, and interfaces.  
A typical software radio consist of a computer and a front end. Embedded software radio platforms also exist, or centralized, cloud-type computing is also possible. 
Table~\ref{Tab:SDR_Compare} compares the features of a variety of SDRs that have been considered for current or future implementation in the AERPAW testbed.

SDRs will be implemented at both BSs and on UASs. At the BS, the SDR must connect to the host system with a high-speed connection, such as 10 gigabit Ethernet. Additionally, these devices are expected to be able to connect to multiple other SDRs over the air. 
Therefore, a high sample rate and MIMO capabilities are also necessary. 
These SDRs may also have larger FPGAs that are useful for offloading some processing. Suitable base stations models include the USRP 2974, N3xx, and X3xx series~\cite{USRP_Website}.

The SDRs carried by the UAVs do not require such high throughput capabilities. These must be lightweight, small sized, and have a lower power demand. Many SDRs that are powered and communicate via USB 3.0 are suitable for a UAV. USB 3.0 is supplied by using mini PCs that are also lightweight, small sized, and have a low power demand. Alternatively, some SDRs come with an ARM processor embedded, allowing for standalone operation. Models that fit the requirements to be used on a UAS include the USRP B2xx and E3xx series and the LimeSDR. It should be noted that these SDRs can also be used at the base station for less resource intensive applications.

\subsection{RF Front Ends}
\begin{figure}
\begin{center}
\includegraphics[width=80mm]{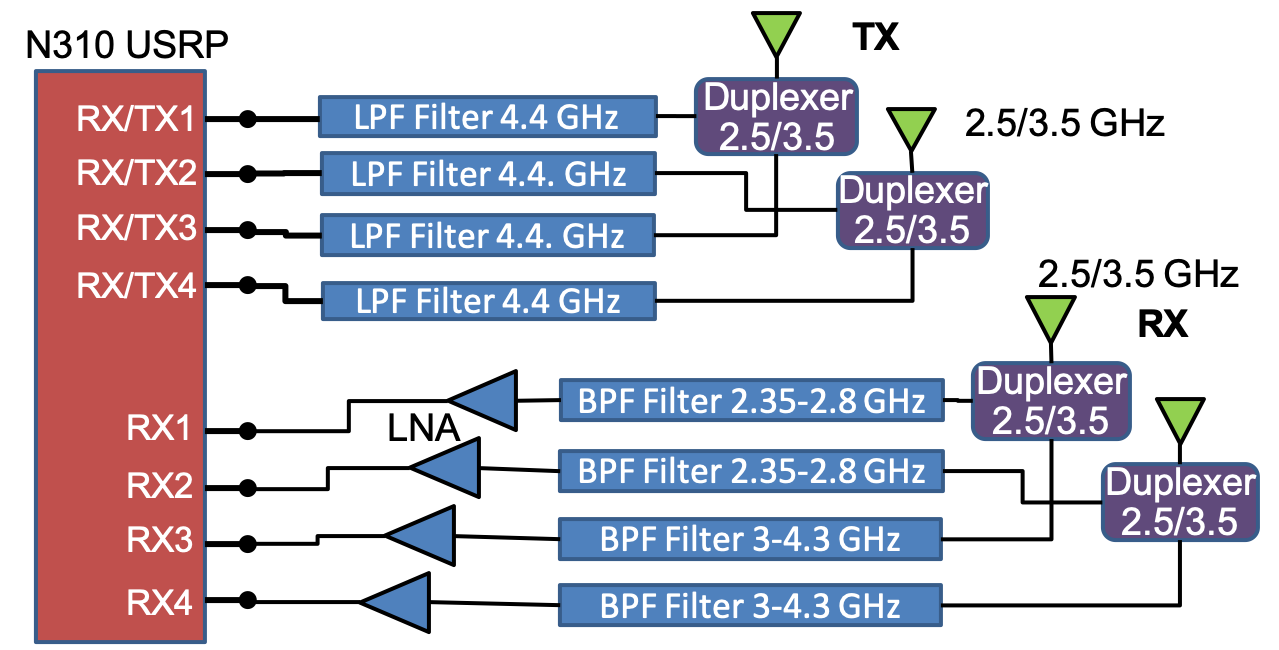}
\end{center}
\vspace{-5mm}
\caption{Proposed RF front end for the Ettus N310 USRP.}
\label{fig:N310}
\vspace{-3mm}
\end{figure}

\begin{figure}
\begin{center}
\includegraphics[width=80mm]{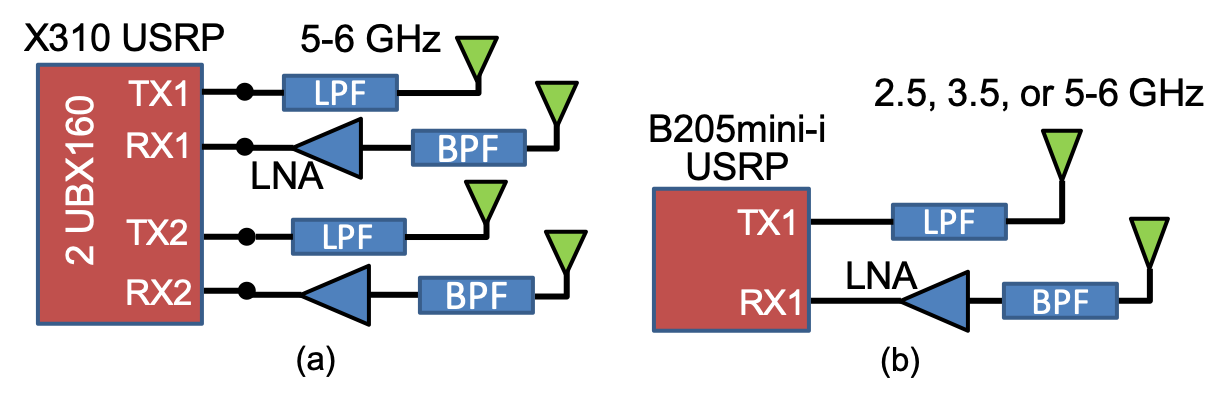}
\end{center}
\vspace{-6mm}
\caption{Proposed RF front end for Ettus X310 (a) and B205mini-i (b) USRPs.}
\label{fig:X-B}
\vspace{-5mm}
\end{figure}

Commercial off-the-shelf SDRs allow a wide frequency range of operation. Because of spectrum regulation and a license needed to access most sub-6 GHz spectrum, initially we will offer support for a limited number of bands. Currently, we are considering the 2.5 GHz, 3.5 and 5.9 GHz bands. Fig.~\ref{fig:N310} and Fig.~\ref{fig:X-B} illustrate our RF architectures as front ends to three USRP models, two used at the tower and one to be carried by typical UAVs. These designs are simple and meant to offer flexibility at low deployment cost and weight.

The proposed antennas for the ground and aerial nodes are RM-WB1-DN-B1K and Octane Wireless SA-1400-5900, respectively, both being wideband and omnidirectional. 

\subsection{SDR Power Measurements}

\begin{figure}
\vspace{-5mm}
\begin{center}
\includegraphics[width=80mm]{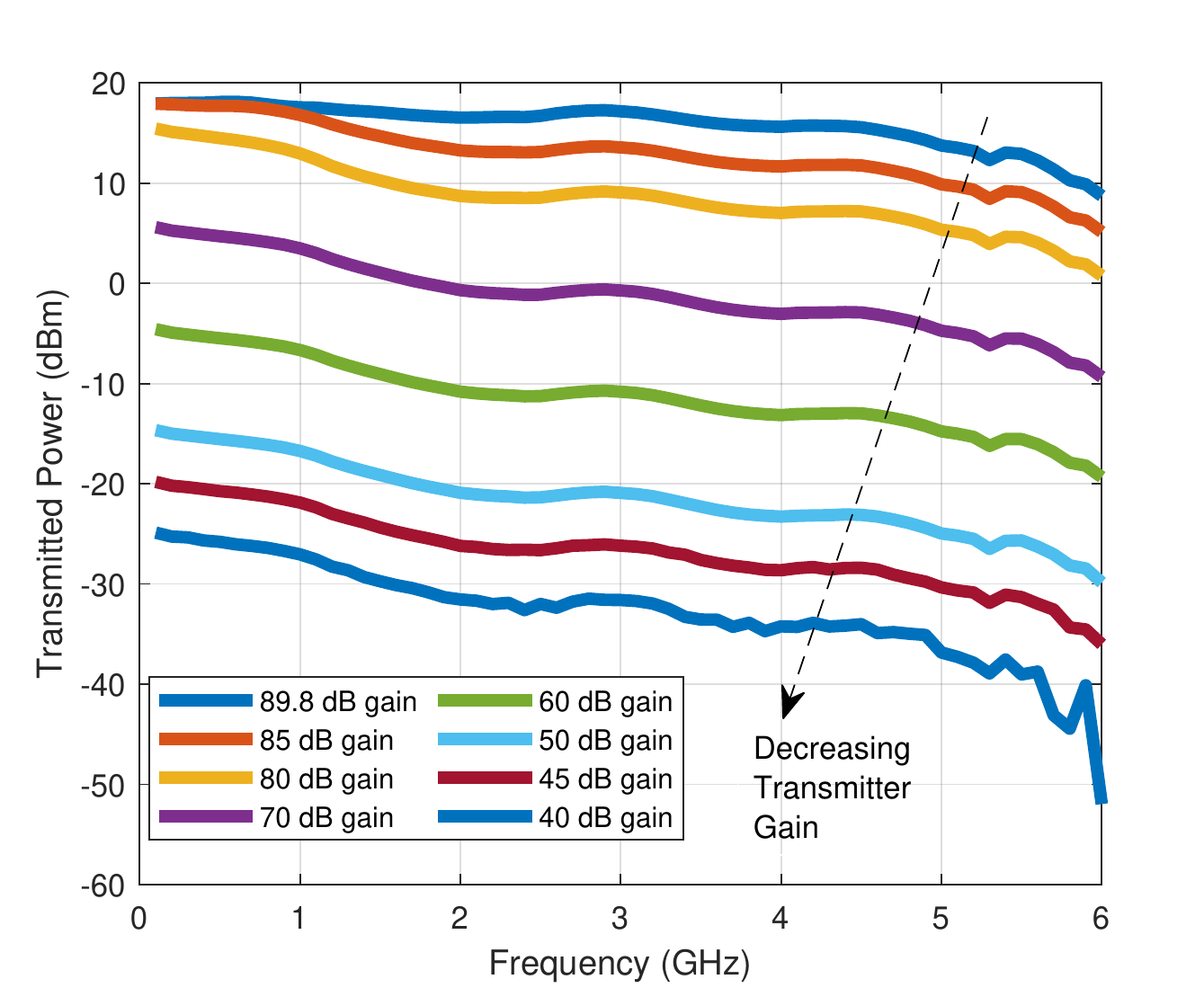}
\end{center}
\vspace{-5mm}
\caption{Transmitted power of a USRP B205mini-i for multiple transmitter
gain and frequencies. 
}
\label{fig:B205gnu}
\vspace{-4mm}
\end{figure}

The output power of the USRP B205mini-i was initially measured by using GNU Radio and a power meter. A second measurement was performed using srsLTE and a spectrum analyzer. In the first measurement, a continuous square wave signal was generated using GNU Radio and measured using a power meter. Fig.~\ref{fig:B205gnu} shows the transmitted power versus frequency for different gain values. The output power saturates as the transmitter gain increases. The results are in close correlation with the values given in the datasheet. 
The data shows that there is an approximately linear relation between transmitter gain and radiated power.  

\begin{figure}
\begin{center}
\includegraphics[width=80mm]{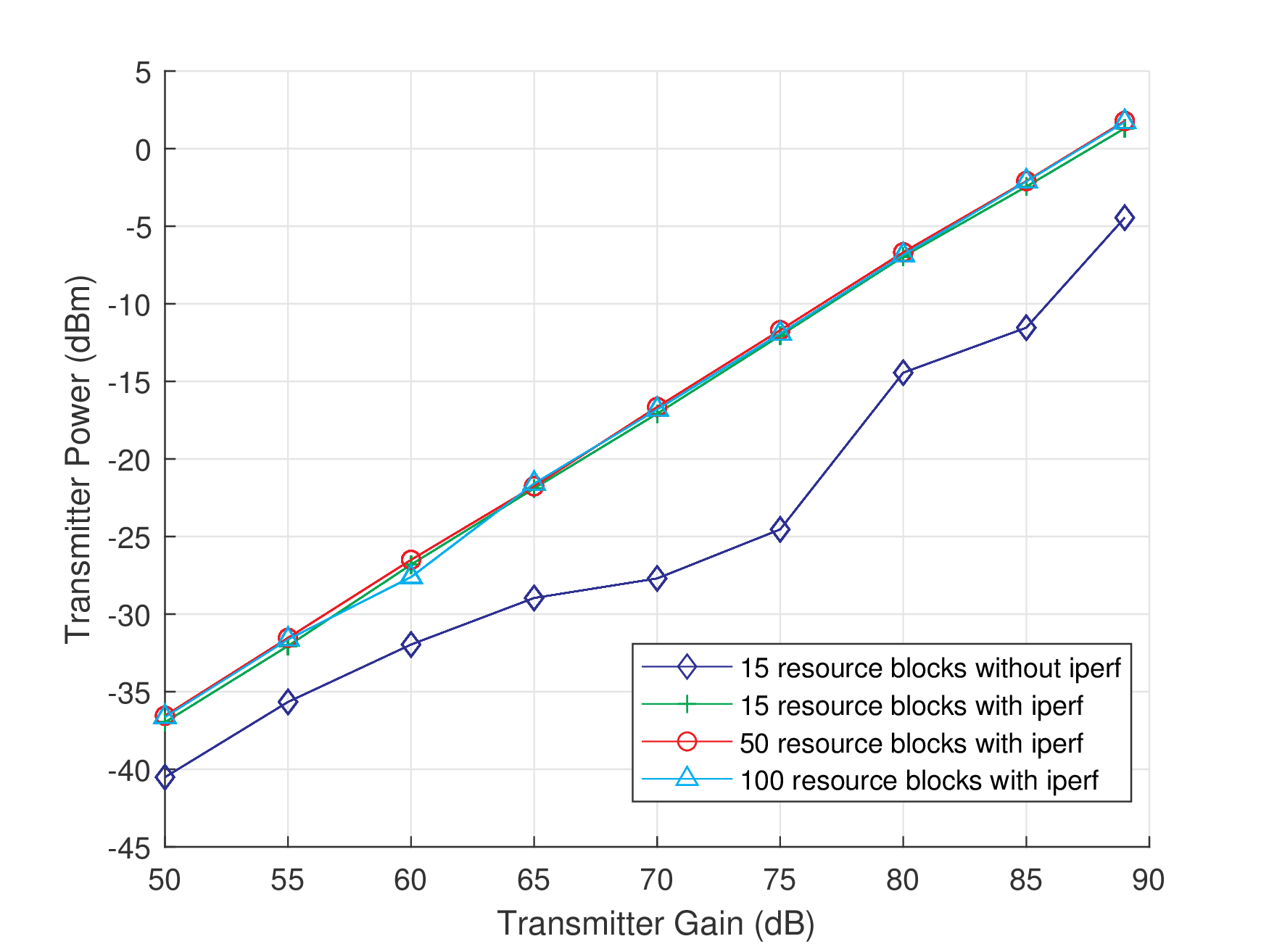}
\end{center}
\vspace{-5mm}
\caption{Plot comparing the maximum transmitted power of a B205mini-i
eNodeB with multiple transmitter gains and resource block quantities. The
transmitted power increased linearly with respect to the gain, except when
data was not transmitted.}
\label{fig:B205powerenb}
\vspace{-3mm}
\end{figure}

\begin{figure}
\begin{center}
\includegraphics[width=80mm]{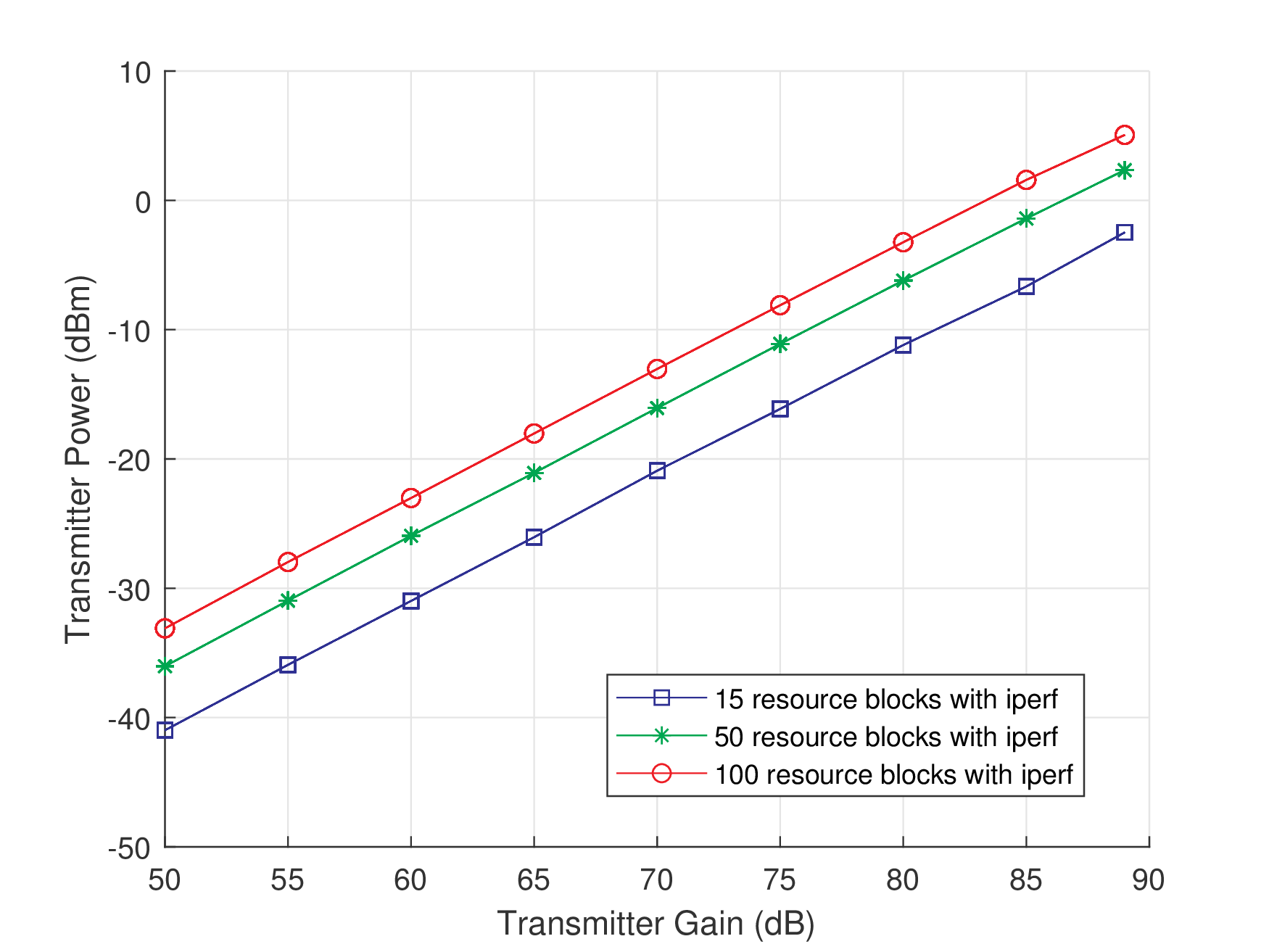}
\end{center}
\vspace{-5mm}
\caption{Plot comparing the maximum transmitted power of a B205mini-i UE
with multiple transmitter gains and resource block quantities.The transmitted
power varies with respect to the number of resource blocks at a constant
transmitter gain.}
\label{fig:B205powerue}
\vspace{-3mm}
\end{figure}

The second measurement was taken using two USRP B205mini-i, with one USRP acting as the transmitter and the other as the receiver. srsLTE and iPerf were used to transmit data using 15, 50, and 100 resource blocks and the power was measured using a spectrum analyzer. The uplink frequency was set to 2565 MHz, and the downlink frequency was set to 2685 MHz. 
Fig.~\ref{fig:B205powerenb} illustrates the variation in the maximum transmitted power of the eNodeB for different transmitter gain settings. All the transmitted signals had a similar gain, regardless of the number of resource blocks. The only case when the transmitted power was lower occured when no data was being transferred using iPerf.
Fig.~\ref{fig:B205powerue} describes the change in the maximum transmitted power of the User Equipment for different gain settings. Here, unlike Fig.~\ref{fig:B205powerenb}, a change in the number of resource blocks produced a change in the transmitted power for the same gain settings.

\subsection{Computers and Interfaces}
The main role of the companion computer is to provide a flexible platform for baseband processing and implementation of the protocol. 
The computers need to be capable of significant processing power and high-speed interfacing. 
To achieve high throughput and real-time operation with the SDR units, an interface such as 10G Ethernet should be used.  Additionally, a low latency kernel can be used to improve real-time operation. 
This can further be improved by utilizing the data plane development kit (DPDK) which can improve the efficiency and throughput of the link when compared to the standard operating system services. 
Our base stations use the Intel Ethernet Converged Network Adapter X520-DA2 in order to take advantage of the N-series and X-series 10G Ethernet capabilities.

\vspace{-0.5mm}

\section{SDR Software}
\label{sec:software}
\subsection{Software Requirements}
The requirements for the software are to be modular, free open-source, robust, and extensible. 
Licensed and closed-source software, such as Matlab and Labview which can interface with SDRs, are also useful for supporting research and will be available through AERPAW. 
Users need to be able to choose, configure, or completely implement the radios, and to facilitate this, experimental tools need to be available with templates and examples.

\begin{figure}
\begin{center}
\includegraphics[width=75mm]{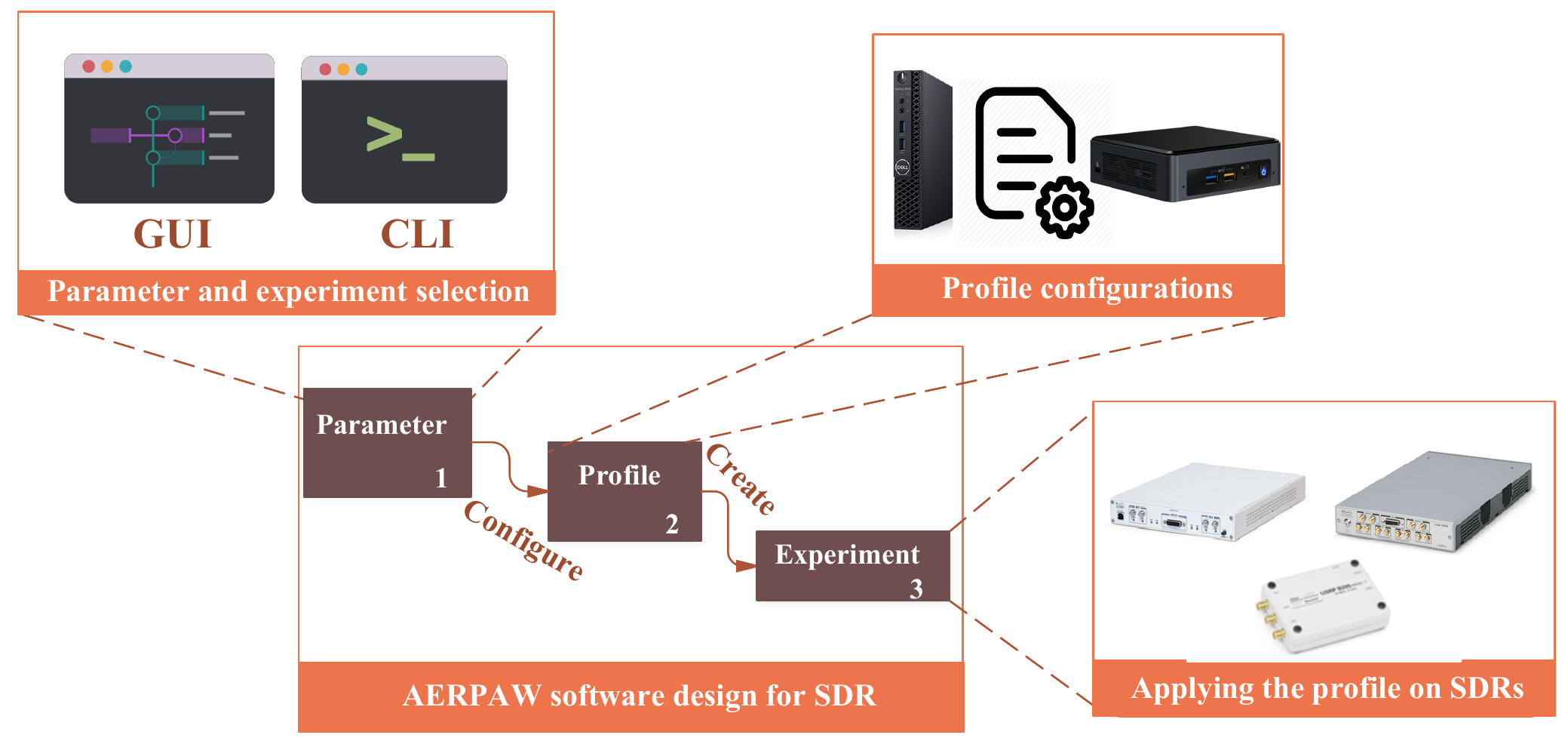}
\end{center}
\vspace{-4mm}
\caption{AERPAW SDR software configuration and experiment flow.}
\label{fig:sdr_flow}\vspace{-5mm}
\end{figure}

\subsection{Waveforms and Protocols}

The SDR software tools will be used in the testbed for providing a stable and reliable environment for research in advanced wireless systems using UAS. Software libraries such as srsLTE and Open Air Interface (OAI) provide LTE and some 5G system features. Both srsLTE and OAI are open-source and allow for the creation of LTE networks using SDRs that can be used by UAS for creating aerial base station, aerial relays, and aerial UEs.
\begin{itemize}
    \item \textbf{srsLTE} is a free and open-source 4G LTE library that can be used to build end-to-end mobile networks using SDRs. The srsLTE library offers the three main components of an LTE system: srsUE implements the user equipment (UE), srsENB implements the LTE base station, or eNodeB, and srsEPC implements the basic LTE core network functionalities, including the MME, HSS, and S/P-GW. The system supports the LTE bandwidths of 1.4, 3, 5 and 10 and 20 MHz~\cite{srslte}.
    \item \textbf{OAI} is a free and open-source library that implements 4G/LTE network with off-the-shelf SDRs. OAI started developing 5G New Radio (NR) in 2017 with a focus on the enhanced mobile broadband (eMBB), using the architecture option 3 of 3GPP, also called EUTRA-NR dual connectivity (EN-DC). The software currently supports a set of 5G-NR physical channels, such as NR-PSS, NR-SSS, NR-PBCH, and NR-PDSCH. The framework uses the FAPI P5 interface to configure the physical layer. The focus is mainly on the development of the NR downlink and frequency range 1 (FR1). 
    The supported sub-carrier spacing is 30 kHz, and the currently supported bandwidths are 40, 80, and 100 MHz~\cite{OAI}.
\end{itemize}


Virtual machines (VMs) will be used in order to prevent users from making modifications to the bare metal system. Software such as srsLTE and OAI will be installed and setup for basic functionality. Hardware connections to the main system will be bridged for access within the VM. 
A user will be able to download the appropriate VMs for their application and run the existing software with little to no modification of the image. 
Both srsLTE and OAI can be run using either two or three VM images, for use with a single UE. The images can be separated into EPC, eNB, and UE, or the EPC and eNB can be joint in a single VM. Running the EPC and eNB on the same image is supported out of the box by srsLTE. OAI allows this type of operation, but does not recommend it. By using VMs over the bare metal machine, approximately 0.1 ms of latency is introduced when running a standard system ping to another system.
AERPAW's SDR design, deployment and experimentation flow is illustrated in Fig.~\ref{fig:sdr_flow}.

\subsection{System Software and Drivers}

The computers will use a Linux setup with VM images that can be easily managed for uploading, downloading, and emulating. We will provide a base image with ``hooks'' allowing 
changes to the default behavior to suit the user's needs. 
The companion computer will not only implement the radio, but also communicate with the autopilot through MAVLink over USB. This will allow controlling the UAS position, orientation and trajectory, 
and receiving information from the navigation sensors. 
Our systems all use Ubuntu 18.04.3 LTS. 
Ubuntu provides a stable platform where all the software of interest is compatible. 
Additionally, its high level of configurability allows for further optimization.





\begin{table}[tb]
\small
\centering
\caption{CPU, memory usage, and downlink throughput of an SDR-based LTE SISO system using srsLTE, where the SmallFF PC is the eNB and the NUC10 is the UE.}
\vspace{-2mm}
\label{tab:my-table2}
\resizebox{0.49\textwidth}{!}{%
\begin{tabular}{|c|c|c|c|c|c|c|}
\hline
\multirow{2}{*}{} &
\multicolumn{2}{c|}{CPU Usage (\%)} & \multicolumn{2}{c|}{Memory Usage (\%)} & \multicolumn{1}{c|}{Downlink Throughput} \\ \cline{2-6} 
                  & SmallFF            & NUC10            & SmallFF             & NUC10             & SmallFF-NUC10          \\ \hline
5 MHz             & 22.0              & 35-45            & 3.9                & 9.5              & 14.7 Mbits/s \\ \hline
10 MHz            & 34.0              & 50-60           & 4.9                & 9.6              & 29.7 Mbits/s \\ \hline
20 MHz            & 70.0              & 100-110          & 6.9                & 10.0             & 59.9 Mbits/s\\ \hline
\end{tabular}%
}\\

\vspace{-5mm}
\end{table}

We use DPDK to further improve the efficiency of the interface with the radio when compared to the standard operating system's interface software. 
This is supported for use with the USRP Hardware Driver (UHD) and allows system cores to be dedicated for input and output operations. 

\subsection{Calibration and Benchmarking}
A range of tests were performed in order to test the average throughput, latency, and stability of open-source SDR software combined with various hardware combinations in both cabled and over the air configurations. 
One SDR benchmarking experiment 
is composed of a B205 mini connected to a Dell small form factor PC as the eNB and a B205 mini connected to an Intel NUC10 as the UE. 
The software used includes srsLTE 19.12, UHD 3.15, and  iPerf3 for LTE throughput measurements. 
Table~\ref{tab:my-table2} summarizes our results.
Throughput tests were also performed using a NUC8 instead of the NUC10. The NUC8 generally had increased CPU usage values at all bandwidths and a very mild decrease in throughput at all bandwidths. All results are average values taken over 5 minute test periods. CPU usage on the UE side was spread over a fair range, therefore, average ranges are listed.
These throughput figures are lower then expected according to the standard, but are reasonably high and scale well with bandwidth.

The NUC8 uses an Intel i7 8559U quad-core processor with 16GB of DDR4 RAM with USB 3.0. The NUC10 uses an Intel i7 10710U hexa-core processor with 16GB of DDR4 RAM with USB3.0. The small form factor PC uses an Intel i9 9900 octo-core processor with 32 GB of DDR4 RAM with the Intel X520-DA2 network interface card, allowing two 10,000 BaseT Ethernet connections. 

\section{Case Study: AERPAW Platform}
\label{sec:services}

\begin{figure}
\begin{center}
\includegraphics[width=70mm]{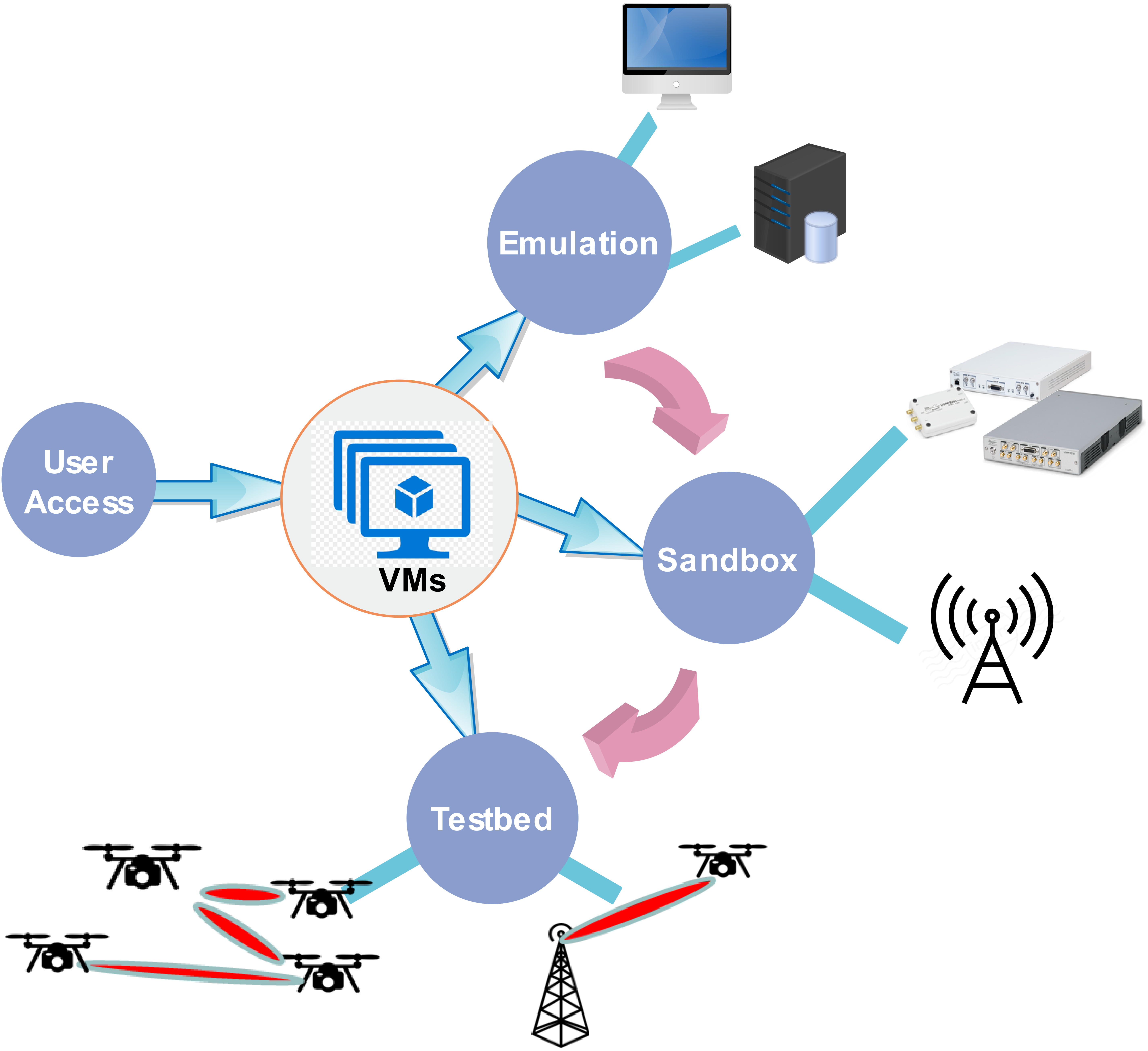}
\end{center}
\vspace{-3mm}
\caption{Overview of the AERPAW resources and experimental flow.}
\label{fig:aerpaw}
\vspace{-5mm}
\end{figure}

AERPAW is a wide-area advanced wireless research testbed with ground and aerial nodes and experimental support tools. Fig.~\ref{fig:aerpaw} depicts a high-level overview of the AERPAW resources and experimental flow.
The testbed will support experimentation with a variety of RF sensors, SDRs, COTS 4G and 5G BSs and UEs, IoT equipment, UWB, and mmWave devices. 
Several radios of each kind will be available to a single experiment 
at deployed testbed locations and portable to other sites. 
Users can be provided with a customized view and front end by using commodity VMs that are connected by virtual local area networks to the UAS and radios assigned to them.

\vspace{-0.5mm}



\subsection{AERPAW Research Requirements}

As part of our AERPAW project, we conducted a survey among researchers and professors. It was determined that full remote access to software-defined radios (SDRs) is a critical feature.  
There was significant interest to use SDRs at cell towers and at the UAS. 
Having access to mmWave SDRs and 5G New Radio (NR) mobile phones carried by the UAS were also highly demanded. 
For UAS applications, search and rescue, detection and tracking, and the use of UAS as hot spot BSs received the highest interest.
Additional experimentation possibilities were suggested by the survey takers. 
These include mobile VR/AR, 
propagation measurements, 
multi-connectivity with UAS, high resolution imaging/surveying, ultra-wideband (UWB),  
computational and energy efficient protocols, 
multi-band and multi-function microwave/mm-wave circuits, extended UAS command and control using cellular, and wireless local area networks \cite{aerpaw_vtc}. 

A flexible and highly efficient experimental research platform is needed to support research on these use cases. 
Moreover, such platform should allow \textit{reproducibility} of experiments, \textit{usability} for quick on-boarding of new users, \textit{interoperability} with new hardware and software, \textit{programmability} of radios and networks, \textit{open access} for spurring innovation, and \textit{diversity} of experiments and users. 
Modern software radio technology, high-density computing, and flexible or plug-and-play RF front ends allow the use of programmable radio systems for development and prototyping, as well as for commercial systems and services.

\subsection{AERPAW Radios}
In addition to the SDRs introduced earlier, AERPAW features a set of commercial radio systems.

\textbf{4G LTE and 5G NR Networks:} Ericsson will provide the 4G and 5G network equipment as a testbed asset. 
Commercial cellular networks will also be used for testbed control. 

\textbf{RF Sensors and Radars:} 
AERPAW will provide RF sensors and radars for research on reliable detection, classification, and tracking of unauthorized UAS. 
\textbf{IoT:} 
LoRa IoT devices will enable experimentation with fixed and aerial nodes. 
Use cases include precision agriculture, where data from the sensors can be collected by the fixed or UAS BSs and processed at the edge or in the Cloud to drive effective decisions at much lower time scales.  

\begin{figure}[t!]
\begin{center}
\includegraphics[width=90mm]{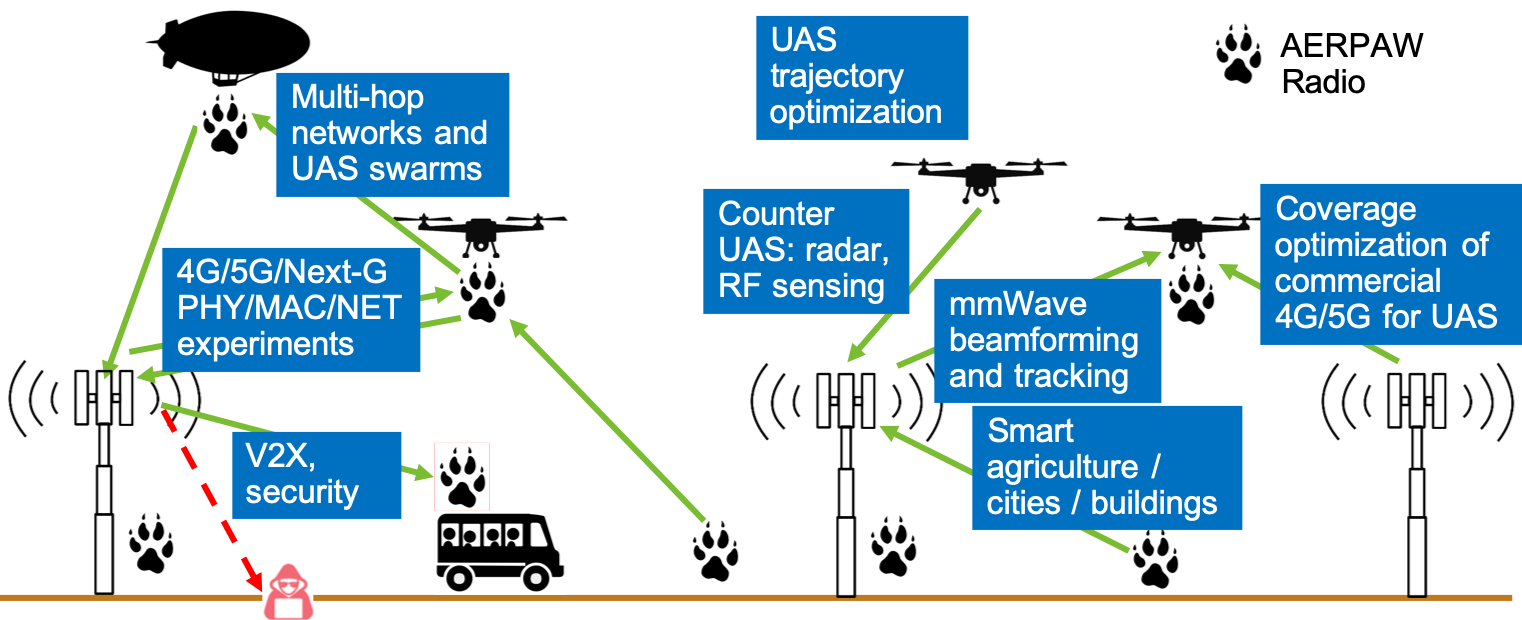}
\end{center}
\vspace{-5mm}
\caption{Overview of the AERPAW management architecture.}
\label{fig:scenarios}\vspace{-5mm}
\end{figure}

\vspace{-0.5mm}

\textbf{UWB and WiFi Sniffers:} 
UWB equipment can support short-range air-to-ground experimentation. 
WiFi sniffers can capture WiFi probe signals from ground nodes, with potential applications in search and rescue, or surveillance. 

\textbf{Millimeter Wave RF:} Custom-built millimeter RF front ends---patch antenna array, beamformer chip and up/downconverter--- for 28 and 39 GHz are envisaged for the UAVs to interface with COTS sub-6 GHz SDRs.

\textbf{Custom Equipment:} Users can also develop and test their own radio software and hardware on AERPAW. 
Custom hardware integration will be subject to size, weight, power, and regulation constraints. 
\subsection{AERPAW UAVs}
AERPAW will have a fleet of custom designed UAVs to have full control of the software and hardware. The UAVs will be of different types: a number of hexacopters and one Helikite. 
The hexacopters can carry 5-10 lbs of radio payload and several of these can form part of a single experiment with appropriate safety precautions (see below).  
Since flight time of these types of UAVs will be limited to 20-40 minutes, depending on payload weight, tethered options will be available for low mobility UAV deployments, e.g. as aerial base stations.
Safety pilots---one per UAV to satisfy the current FAA regulations---will be available for the experiment.

\subsection{Experiment Development}
A node computer on a UAV or at a tower
will run a VM whose image is under the control of the experimenter. 
The experimenters will be offered example radio profile images  
for implementing different networks or network elements in software and testing it in the emulator/sandbox, before testbed deployment (Fig.~\ref{fig:aerpaw}). 
Each VM in the experiment can be configured in two ways: the user can either directly configure a clone of a template VM residing in the experimenter space and then use the resulting VM in the experiment, or download an image template of the VM, configure it locally, create a new image and send it back for testing. 

Experiment and code verification will be done by the user, using the AERPAW emulator and sandbox. 
The main advantage of emulation as opposed to simulation, 
is that the same VMs can run on the emulated system, sandbox 
and eventully on the real testbed. 
Once an experiment is verified, the user can submit the configuration files for deployment, specifying the AERPAW sites and the physical nodes that the experiment needs. 
The execution of certain experiments can be fully automated, which is the ultimate goal. 
Fig.~\ref{fig:scenarios} illustrates AERPAW testbed nodes and sample research  experiments.


\section{Conclusions}
\label{sec:conclusions}
This paper presents our approach to radio system development and testing for supporting research and future UAS platforms. We leverage SDR technology as the foundations for building a new advanced wireless testbed for aerial scenarios: AERPAW. 
The proposed hardware design and open-source software are to be used at  fixed radios deployed on towers,  and portable radios to be carried by small UAVs. 
The open nature of our modular design with exchangeable COTS components allow easy reconfiguration and portability of radios to other platforms, such as other aerial or ground vehicles or test facilities. 
Extensive 3D vehicle and wireless emulation as well as real-world testing with varying UAV heights, radio environments, and antenna configuration are the next steps for AERPAW. Our future work also includes software development to support robust 5G waveforms and protocols on the platform.


\section*{Acknowledgment}
This work was in part supported by the NSF PAWR program under grant number CNS-1939334. 


\balance

\bibliographystyle{IEEEtran}
\bibliography{imsi_bib,vuk,CAREER_isma,ismabib}

\end{document}